\documentstyle[12pt,titlepage]{article}
\def \p {\mbox{\boldmath $p$}}

\begin{document}
\begin{titlepage}
\vspace*{0.5truein}
\begin{center}
{\Large\bf SCALAR THREE BODY DECAYS AND SIGNALS FOR NEW PHYSICS\\[0.5truein]}
{\large Rathin Adhikari$^1$\\}
Mehta Research Institute, \\
10 Kasturba Gandhi Marg,  Allahabad 211 002, INDIA \\
{\large Biswarup Mukhopadhyaya$^{2,*}$\\}
International Centre for Theoretical Physics \\
Trieste, ITALY \\
\end{center}
\vskip .5cm

\centerline{\bf ABSTRACT}
If massive invisible particles are pair-produced in a three-body decay,
then the energy distribution of the other (visible) product is sensitive to
the mass of the invisible pair. We use this fact in the contexts of a
Higgs boson
decaying into (i) a $Z$-boson and two massive neutrinos of a fourth
generation, and (ii) a Z and two lightest supersymmetric particles in the
minimal supersymmetric standard model. We discuss how the Z-energy
spectrum in each case can reflect the values of the parameters of such
models.

\hspace*{\fill}

\noindent
PACS number(s): 13.85.Qk., 14.80.Gt., 14.80. Cp. \\
$^{1}E-mail : ~~rathin@mri.ernet.in  \;\; ^{2}E-mail :
biswarup@mri.ernet.in$

\noindent
$^{*}Permanent \; Address: \; Mehta \; Research \; Institute,
 \; 10 \; Kasturba \; Gandhi \; Marg, \; Allahabad-211 002, \; INDIA$

\end{titlepage}

\textheight=8.9in

It is a known fact that if a pair of invisible particles are produced in
some three-body decay, then the energy distribution of the third
particle is sensitive to the masses of the invisible particles. This
sensitivity
has been utilised earlier in the context of rare decays like
$K^+{\longrightarrow} \pi^+ \nu \bar{\nu}$ to study the dependence of the
resulting pion spectrum on the mass of the $\tau$-neutruino \cite{Desh}.
Also, it
has been claimed that the decay spectra in such cases are different for
Dirac and Majorana neutrinos respectively, thereby suggesting a method for
distinguishing between these two kinds of fermion masses \cite{Pal}.

The essential argument in the above works is as follows. If all the
neutrinos have masses that are negligible compared to $m_\pi$, then
the differential decay rate $d\Gamma / dE_\pi$ in the centre-of-mass
frame will be a monotonically
increasing function of $E_\pi$ over the allowed region of phase space, as
can be seen from straightforward kinematics. If, on the other hand, one of
the
neutrino species is significantly massive, then the decay distribution for the
corresponding channel attains a peak and then falls with increasing $E_\pi$,
due to the unavailability of phase space. As a result,
$ \sum_{i}{{d\Gamma_{i}}\over{dE_{\pi}}} $
exhibits a kink (i is the generation label). With increasingly higher mass
of the invisible pair, the
kink is displaced progressively to lower energy regions. However, as
higher mass implies more phase space suppression for the channel under
question, the consequent distortion in the decay spectrum also tends to be
less and less conspicuous. In between, there is an optimal region where
one expects the highest sensivity to the mass of the invisible pair.

Since the current upper bound on the $\tau$-neutrino mass from laboratory
measurements \cite{Data1,Data2} is as low as $35 MeV$, the idea summarised
above
is of little potential use in its original context; the kink in the
$\pi$-spectrum in $K^+ {\longrightarrow} \pi^+ \nu \bar{\nu}$ can barely
occur
at the very edge of the phase space even if $\nu_\tau$ has a mass close to
its upper limit. However, because of its essentially kinematic origin, a
similar effect in the decay distributions of heavier particles can also be
expected. This should have interesting applications in obtaining the
signatures of new particles which may be invisible in character. As an
example, we consider in this note the decay channels of a Higgs boson into
a $Z$ and two invisible fermions, and show how the decay spectra are
sensitive to the masses of such fermions. This confirms the expectation
of a kink-like behaviour even when the visible decay product is a particle
with spin (whereas the original observation concerned only spinless mesons).
We illustrate such behaviour in the contexts of two types of decays,
namely (a)$H{\longrightarrow} Z N \bar{N}$ where $N$ is a heavy sequential
Dirac neutrino \cite{fourth},
and (b)$H {\longrightarrow} Z \chi^0 \chi^0$ where $\chi^0$ is the lightest
supersymmetric particle (LSP) in the supersymmetric (SUSY)
extension of the standard model \cite{Review,Report}.

Of course, we are conscious of the fact that the decays of the Higgs boson
into the above channels will have rather small branching ratios, and that,
considering the unavoidable backgrounds in hadron colliders, the
observation of decay patterns of the expected type poses practical problems.
Still we find it worthwhile to undertake this study because of two main
reasons.
First, as we have already mentioned, it enables one to see the effects in a
general perspective, whereby some insight might be gained about the signals
of massive invisible particles in theories beyond the standard model. In
addition, such decay
distributions could be useful as cross-checks of the signatures of invisible
particles whose usual search strategy is to look for missing $p_T$
\cite{Collider}. Such
cross-checks are particularly helpful if more than one types of
non-standard physics show up at the same time in experiments.

As the first example, we consider the decay $H{\longrightarrow}Z N \bar{N}$,
where N is a heavy neutrino belonging to, say, a fourth generation
\cite{fourth}. As
we know from the measurement of the $Z$-width \cite{Data2}, $m_N > m_Z/2$.
If such
a neutrino exists and happens to be lighter than its corresponding
charged lepton, then its only possible mode of decay is into a $\tau$ and a
real or a virtual $W$. But the decay is controlled by the mixing angle
between the third and fourth generations. From a purely phenomenological
standpoint it is possible to have a very small value ($10^{-8}$ or less)
of this mixing angle. Under such circumstances, the decay length of the
heavy neutrino can be so large that it may pass off as invisible in a
hadronic collider. The final state in the mode considered above will then
essentially consist of $Z +\not{P_T}$, similar in nature to the situation
where massless, standard model neutrinos are pair-produced along with the Z.

Assuming standard model couplings and keeping the mass of the neutrino, the
differential decay width for $H{\longrightarrow} Z \nu_i \bar{\nu_i}$
in the rest frame of the decaying $H$ is given by

\hspace*{\fill}

\noindent
{\large ${{d\Gamma_{i}}\over{dE_{Z}}} =$  }
$  \; \; {\left( \mbox{ \large ${{g_{w}^4 \;
\lambda_{i}}
\over {64 \;m_{Z}^{4} \;m_{H} \;\cos\theta_{w}^{4}\;\pi^{3} }}$ } \right) }
\; \mbox{\large {\bf [[ } $  { \; 1 \over {{ ( \;m_{H}^2-2
\; m_{H}  \;
\;E_{Z} )}^2
+ \;m_{Z}^2  \;  \Gamma_Z^2 }} $ {\bf \{ } } \; 2 \; m_{Z}^8
 +
\;m_{Z}^6
\;m_{H}^2 \\ -  \;m_{Z}^4 \;  \;m_{H}^4 -2
\;  \;m_{\nu_{i}}^2  \;  \;m_{Z}^2  \;  \;m_{H}^4
-2 \;m_{Z}^6 \;m_{\nu_{i}}^2+2 \;E_{Z} \;m_{Z}^2 \;m_{H}  \; ( -2 \;m_{Z}^4
+2
\;m_{H}^2 \;m_{\nu_{i}}^2)  \\ -2 \;E_{Z}^2 \;m_{H}^2 \; ( \;m_{Z}^2
\;m_{\nu_{i}}^2 +3
\;m_{H}^2 \;m_{\nu_{i}}^2)+4 \;E_{Z}^3 \;m_{H}^3 \;m_{\nu_{i}}^2+ \;m_{Z}^4 \;
m_{H}^4+ \;m_{Z}^4 \;m_{H}^2 \;E_{Z}^2 - \\
\;m_{Z}^4 \;
m_{H}^2 \;\lambda_{i}^2 \;/3
+2 \;m_{H}^5 \;E_{Z} \;m_{\nu_{i}}^2+4 \;m_{Z}^4 \;m_{H} \;E_{Z}
\;m_{\nu_{i}}^2
+ \;m_{\nu_{i}}^2 \;( \;m_{H}-2 \;E_{Z}) \;[ \;m_{Z}^4 \;m_{H}^2-4
\;m_{Z}^4 \;E_{Z}
\;m_{H} \\ + \;m_{H}^4 \;E_{Z}^2+2 \;m_{Z}^6
-2 \;m_{Z}^4 \;m_{\nu_{i}}^2  +2 \;m_{Z}^2 \;m_{\nu_{i}}^2 \;m_{H}^2-2
\;m_{H}^2 \;
m_{\nu_{i}}^2 \;E_{Z}^2- \;m_{H}^4 \;E_{Z}^2 \;]
\;(\omega_{i}/\lambda_{i}) \;
\mbox{\large {\bf \} } }  \;+ \\    \mbox{ \large ${ \;m_{\nu_{i}}^2 \over {
\;m_{Z}^2+4 \;
m_{\nu_{i}}^2 \; {\boldmath{p}}_{Z}^2 /q^2}} $}
\;[  \; - \;m_{Z}^4 + \;m_{Z}^2 \;m_{\nu_{i}}^2+ \; 2 \;( \;m_{Z}^4
\;
m_{\nu_{i}}^2 - \;m_{Z}^2 \;m_{\nu_{i}}^4 \;)/ \;m_{H}^2  \; ] \\
+  \;m_{\nu_{i}}^2 \;\omega_{i} \; [ \;m_{Z}^2 \;E_{Z} - 2 \;m_{Z}^2 \;
m_{\nu_{i}}^2/ \;m_{H} \;
\; ] /\;\lambda_{i} \;+ m_{\nu_{i}}^2 \;(- \;m_{Z}^2+ \;m_{H}^2-2
\;m_{H} \;
E_{Z}-2 \;m_{\nu_{i}}^2  \; )/2  \; \; \mbox{\large {\bf ]]  } }
$

\hfill (1)

\medskip
\noindent
with\\
$
q^{2}=\;m_{Z}^2+\;m_{H}^2-2 \;m_{H}\; E_{Z}
$
\hfill (2a)

\noindent
$
\p_{Z}^2=\;E_{Z}^2-\;m_{Z}^2
$
\hfill (2b)

\noindent
$
\lambda_{i} =\p_{Z} {(1 - 4 \;m_{\nu_{i}}^{2}/\;q^{2})}^{1\over 2}
$
\hfill (2c)

\noindent
$
\omega_{i} = \ln$ {\large $ \left({{E_{Z}+\;\lambda_{i}}\over
{E_{Z}-\;\lambda_{i}}}\right)$}
\hfill (2d)

\hspace*{\fill}

\noindent
where $i$ is the generation label, $E_Z$ is the $Z$ energy in the rest
frame of decaying Higgs and
the kinemetical constraint on $Z$ -energy is

\noindent
$m_Z \leq E_Z \leq (m_H^2 +m_Z^2 - 4 m_{\nu_i}^2)/2 m_H
$
\hfill (3)

\hspace*{\fill}

The contributions come from three Feynman graphs, there being a
non-negligible $H \bar{\nu} \nu$ interaction if $\nu$ is massive. For
$i=1-3$, we obtain
the appropriately simplified expression by setting $m_{\nu_i} =0$. With a
massive neutrino, i runs from 1 to 4, with $\nu_4 = N$. The differential
rate for the entire $Z + invisible$ channel is obtained by adding the
contributions from the massless as well as massive species.

Figure 1 illustrates the patterns expected for the decay of a Higgs of
mass $500 GeV$, with three massses for the heavy neutrino. The kink due
to the presence of the heavy species begins to become perceptible as
$m_N$ approaches $100 GeV$. With higher $m_N$, the kink appears, as
expected, for lesser values of $E_Z$. However, in contrast with the results
presented in reference \cite{Desh}, the distortion in the curve for $m_N =
150 GeV$
is more pronounced than that for $m_N = 100 GeV$. This is due to the fact
that unlike in the case of meson decays, here we have a resonant
contribution to $H
{\longrightarrow}
Z \nu_i \bar{\nu_i}, i=1-3$ from real Z-bosons. This resonance
(not shown in the figures) occurs at
$E_Z = m_H/2$. The kink for $m_N = 100
GeV$ occurs sufficiently near the
resonance to be partially washed away by the sharp rise that ensues. That
is why the fall for $m_N = 150 GeV$  is somewhat more marked, although
for even higher masses the kink again starts losing visibility.

Let us now turn our attention to the minimal SUSY standard model where
invisible Higgs decay mode
\cite{hunter,invisible} could offer signal for new physics. We consider
here a decay mode
$H {\longrightarrow} Z + invisible$ which can occur in a less restricted
region of the parameter space. The invisible particle in this case is the
lightest supersymmetric particle which can decay no further because of
the conservation of R-parity \cite{Rparity}. In most theories, the
favoured candidate
for LSP is the lightest of the neutralinos, the physical states obtained
upon diagonalisation of the mass matrix consisting of the photino, the Zino
and two neutral Higgsinos \cite{Report,Higgs rule}. As is well-known, this
scenario contains two
complex scalar doublets which ultimately give rise to two neutral scalars
(along with one pseudoscalar). Of these, the upper limit on the mass of the
lighter scalar is a function of the top quark mass \cite{Radiative}, and
cannot be
appreciably
above $150 GeV$ or so. Under the circumstances, its decay into
$Z \chi^0 \chi^0$ ($\chi^0$ being the LSP) does not have much leeway
kinematically. The heavier scalar, on the other hand, is more suitable for
demonstrating the decay behaviour we are interested in. In what follows
we present the sensitivity to the $\chi^0$-mass in the differential decay
rate of this mode plotted against $Z$-energy. It may perhaps be said that
if more than one scalar particles are experimentally discovered while direct
evidences for SUSY (from squark and gluino production) are still not
available, then such indirect signals through scalar decays may be useful
in resolving the issue in favour of SUSY or otherwise.

The actual computaion of the decay rate for $H {\longrightarrow}Z \chi^0
\chi^0$  involves evaluation of the contributions from ten Feynman diagrams,
having as propagators the four neutralinos (with their crossed
diagrams), the Z and the pseudoscalar particle and the corresponding
Feynman rules are given in references \cite{Report,hunter,Higgs rule}.
The large number of parameters involved can be simplified
if  one considers a SUSY scenario inspired by Grand Unified Theories
(GUT) \cite{GUT}. There, all the neutralino masses and their mixing angles
(which
occur in the couplings that we require to know) can be obtained using as
inputs the gluino mass ($m_{\widetilde{g}}$), $\tan \beta =v_2/v_1$
where $v_2(v_1)$ is the vacuum expectation value of the Higgs doublet
that gives mass to the up(down)-type quarks, and $\mu$, the Higgsino
mass parameter \cite{Para1}. Also, in the Higgs sector in the minimal SUSY
model,
$\tan \beta$ and the mass of one physical scalar suffice to fix all masses
as well as $\alpha$, the mixing angle between the two doublets. As a net
outcome, it is possible to compute the decay rates mentioned above by
specifying $m_{\widetilde{g}}$, $\tan \beta$, $\mu$ and the mass of H, the
decaying particle. A further constraint on the allowed combination of
these parametres comes from Z-decay measurements at the Large Electron
Positron (LEP) collider \cite{Para2}.

In addition to the SUSY contribution, one has also to add
the contributions from the three massless neutrinos in order to have the
net observable distributions of the $Z + invisible$ final states. The
neutrino contribution is given by equation (1), with $i=1 - 3 ,
m_{\nu_{i}}=0 $ and an overall
multiplicative factor of $ \cos^2 (\beta - \alpha)$. The expression for the
SUSY contribution being extremely long and cumbersome, we refrain from
presenting it here.

We present some of our results in figures 2 and 3, drawn for similar values
of all other parameters but with opposite signs for $\mu$. The choice
of a Higgs mass of $500 GeV$ is because in this region (upto an interval
of about $50 GeV$) the distortion to the decay spectrum is somewhat
optimal. A large
fluctuation due to the LSP is observed in each curve which otherwise
would have had a uniform rise due to the neutrino contributions alone. The
several orders of enhancement caused by the neutralinos over the neutrinos
can be understood from the fact that for the massless neutrinos, the only
contributions to this mode can be mediated by a transverse Z-boson. For both
the figures here we use the value $ \tan \beta = 2$; for larger values of
$\tan \beta$ the fall is sharper but, owing to a stronger suppression
caused by the factor $ \cos^2 (\beta -\alpha)$ in this case, the subsequently
rising part from the neutrino contributions is extremely small. Also, for
a lower $m_H$, the SUSY contribution tends to become smaller compared to
the standard model one, so that the distortion begins to disappear.

It is to be noted that in the SUSY case we are in a kinematic region where
the decay
spectrum is not plagued by resonances, so that a clean signature of the mass
of the LSP can be observed. Here one can see a close resemblance to the
curves of refenence \cite{Desh}, the kink (which in this case is trully a
``hump'') being progressively in the region of smaller $E_Z$ and at the same
time being smaller in size as the invisible superparticle is more massive.

The production cross-section of a $500 GeV$ Higgs at the LHC is about
$2-3 pb$ \cite{lhc}. This means that about $10^{6}$ Higgs bosons can be
produced per year. Since the branching ratios for the three-body  decays
under question are on the order of $10^{-4}$, a few hundreds of such events
in a year of run are possible. In the SUSY scenario, as one see from figures
2 and 3, a large contribution to the observed events are expected from that
part
of the Z-energy spectrum where the distortion due to the massive LSP is
visible. This makes the SUSY scenario more interesting for
demonstrating the effects we are trying to show and highlights the
possibility of
indeed observing the signals of the massive LSP's in this manner. In the case
of massive neutrinos, the number of events in the distorted region are
possibly too small to make the effect experimentally interesting.

We conclude by re-iterating that the predictions made here are mainly aimed
to  bring forth
into a bigger perspective the issue of obtaining signals of invisible
particles
from the observed energy spectra of visible ones. Detailed studies on some
possible applications to B-factories as well as to the LEP-II are
currently in progress \cite{rb}.

\hspace*{\fill}

\hspace*{\fill}

\noindent
{\large {\bf Acknowledgment}} \\

R.A. would like to thank Professor H. S. Mani for helpful discussion
and comments. B. M. acknowledges helpful comments from A. Smirnov, and
also the hospitality of the International
Centre for Theoretical Physics, Trieste, where this work was completed.

\newpage
\centerline {\large {\bf Figure Captions}}

\hspace*{\fill}

\hspace*{\fill}

Figure 1: \\

The differential decay distributions for $H {\longrightarrow} Z + nothing$
in the presence of a massive but invisible fourth neutrino $N$. The three
curves correspond to different values of $M_N (in GeV)$.

\hspace*{\fill}

\hspace*{\fill}

Figure 2: \\

The differential decay distributions (modulo an overall multiplicative
factor) for $H {\longrightarrow} Z +
nothing$ in the minimal supersymmetric standard model. The three curves
correspond to different masses of the LSP $(in GeV)$, with $\mu = 250
GeV$, $tan \beta = 2$.

\hspace*{\fill}

\hspace*{\fill}

Figure 3: \\

The differential decay distributions (modulo an overall multiplicative
factor) for $H {\longrightarrow} Z + nothing$
in the minimal supersymmetric standard model. The three curves correspond
to different masses of the LSP $(in GeV)$, with $\mu = -250 GeV$,
$tan \beta = 2$.

\end{document}